\documentclass[useAMS,usenatbib]{mn2e}

\usepackage{graphicx}

\title[On the orbit orientation of $\eta$ Carinae]
{Constraining the orbital orientation of $\eta$ Carinae 
from H Paschen lines}
\author[D. Falceta-Gon\c{c}alves \& Z. Abraham]
{D. Falceta-Gon\c{c}alves$^{1}$\thanks{e-mail:diego.goncalves@cruzeirodosul.edu.br} 
and Z. Abraham$^{2}$\\
$^{1}$N\' ucleo de Astrof\' isica Te\' orica, Universidade Cruzeiro do 
Sul, Rua Galv\~ ao Bueno 868, 01506-000, S\~ao Paulo, Brazil \\
$^{2}$Instituto de Astronomia, Geof\'\i sica e Ci\^encias Atmosf\'ericas, Universidade de 
S\~ao Paulo - Rua do Mat\~ao 1226, 05508-090, S\~ao Paulo, Brazil}

\begin{document}
\date{ }

\pagerange{\pageref{firstpage}--\pageref{lastpage}} \pubyear{2007}

\maketitle

\label{firstpage}

\begin{abstract}
 During the past decade several observational and 
theoretical works provided evidences of the binary nature of $\eta$ Carinae. Nevertheless, 
there is still no direct determination of the orbital parameters, and the different current 
models give contradictory results. The orbit is, in general, assumed to coincide with the Homunculus 
equator although the observations are not conclusive. Among all systems, $\eta$ Carinae 
has the advantage that it is possible to observe both the direct emission of line 
transitions in the central source and its reflection by the Homunculus, which are 
dependent on the orbital inclination. In this work, we studied the orbital phase 
dependent hydrogen Paschen spectra 
reflected by the SE lobe of the Homunculus to constrain the orbital parameters of 
$\eta$Car and determine its inclination with respect to the Homunculus axis. 
Assuming that the emission excess is originated 
in the wind-wind shock region we were able to model the latitude dependence of the 
spectral line profiles. For the first time, we were able to
estimate the orbital inclination of $\eta$Car 
with respect to the observer and to the Homunculus axis. The best fit occurs for an orbital inclination to the line of sight of 
 $i \sim 60^{\circ} \pm 10 ^{\circ}$, and  $i^* \sim 
35^{\circ} \pm 10 ^{\circ}$ with respect to the Homunculus axis, indicating that the angular 
momenta of the central object and  the orbit are not aligned.  We were also able to fix the phase angle of conjunction as 
$\sim -40^{\circ}$, showing that the periastron passage occurs shortly after conjunction.
\end{abstract}


\section{Introduction}

Among the most massive and luminous stars in the Galaxy $\eta$ Carinae is possibly the most enigmatic and intriguing 
object. Unique features, e.g. variable emissions and the outbursts that occured during 
the 1800's, which generated the Homunculus - the giant 
bipolar gas structure surrounding the star -, made difficult the understanding of the 
real nature of this object. Its periodicity 
in the observed light curves and spectral lines at all wavelengths indicated its binary 
nature with $P = 5.52$ years (Damineli 1996, Corcoran et al. 2001, Stahl et al. 2005). 

From the X-ray light curve and spectra it is possible to infer the 
existence of a strong wind-wind shock, expected for an O-type or Wolf-Rayet (WR) 
companion star (Pittard \& Corcoran 2002). However, since there is no direct observation 
of the individual stars of the system it is difficult to precisely derive its 
orbital parameters, e.g. the eccentricity, inclination and periastron position, and indirect methods have to be used. 
The periodic variations in the X-ray light curve, which is originated by the wind-wind 
shocked gas during periastron passage, has already provided some clues on this issue 
(Corcoran 2005). 
Similarly, the sharp drop in flux  has also been observed at 
optical (Fern\'andes-Laj\'us et al. 2008), infrared (Whitelock et al. 2004) and radio 
wavelengths (Duncan \& White 2003, Abraham et 
al. 2005a), but with different duration. Corcoran et al. (2001) assumed that the "eclipse-like" dips are produced by increased 
absorption in the stellar envelope. In this scenario, the orbital 
plane should closely intercept the line of sight and the periastron should be located in 
opposition regarding the observer. 
Falceta-Gon\c calves, Jatenco-Pereira \& Abraham (2005) proposed a different scenario, in 
which the remnant post-shocked gas could increase the optical depth and 
explain the observed X-ray light curve if the periastron passage is close to conjunction. In a further work, it was also possible to 
reproduce the radio light curves under this assumption (Abraham et al. 2005b). The same 
process could be invoked to explain the decrease in optical and infrared fluxes.

Although X-rays, optical, infrared, and radio light curves may provide clues on the 
orbit of the system, and even be used on the accurate estimation of the stellar wind 
properties, it is well known that modelling spectral line profiles is the best method to 
determine orbital parameters of binary systems, in general, with good precision. 

Damineli (1996), studying the HeI$\lambda 10830$, 
reported a periodicity of 5.5yrs in the spectroscopic events, the epochs in which the 
line disappear. Steiner \& Damineli (2004) studying the higher excitation energy lines 
of He II, on the other hand, reported a peak in intensity right before periastron 
passage. Compared to the X-ray light curve, the similarity reveals that they may have the 
same origin. Further studies, with higher resolution 
spectra (Martin et al. 2006), have shown that, not 
only the line intensities vary within the 5.5yrs period, but also their profiles.
They concluded that the line source might be more 
complex than simply atmospheric emission, and the decline in flux cannot be explained only by 
eclipsing models. 
Nielsen et al. (2007) attempted to explain the line-profiles  
assuming that the emission is the combination of a major component originated 
in the stellar wind, and other minor contributions at different velocities, which origin is not clear. From the 
line velocity curve, they inferred an eccentricity $e \sim 0.9$, and that  periastron 
should be placed near opposition. 

Falceta-Gon\c calves, Jatenco-Pereira \& Abraham (2007) showed that line profile 
variability is expected from emission originated in the wind-wind interaction zone. Its study showed to be a 
powerful tool in determining the orbital parameters of WR30a (Falceta-Gon\c calves, 
Abraham \& Jatenco-Pereira 2008), where by fitting synthetic profiles to the observational
data,  it was possible to determine the 
inclination and eccentricity of the orbit. Also, using the line velocities obtained 
directly from the primary O-star,  it was possible to constrain the stellar 
masses. 

In the $\eta$ Carinae binary system, observations of direct and Homunculus reflected emission show multi-peaked 
spectral lines, with profiles and intensities that vary with orbital phase. Stahl et al. 
(2005) showed that the HeII $\lambda4686$\AA \ emission lines measured towards the 
central object and the SE lobe of the Homunculus presented different profiles, in 
contradiction to what would be expected assuming spherical symmetry. 
Abraham \& Falceta-Gon\c calves (2007) showed that the conical wind-wind shock region 
could be the main source of the observed excess in the HeII $\lambda4686$\AA \ line 
profiles.
 To reproduce both the direct and Homunculus reflected line profiles it was necessary to assume 
that the Homunculus axis does not coincide with the axis of the orbital plane. However, 
due to the limited amount of data, which were restricted to orbital phases very close to periastron passage, 
it was not possible to lift the degeneracy regarding the inclination of the orbital and  
Homunculus axis. In all cases, the observed velocity curve, as well as the line profiles, are well reproduced 
only if  periastron occurs near conjunction.

In the present work we extended the analysis of Abraham \& Falceta-Gon\c calves (2007), 
applying this model to the H Paschen 8 line  reflected 
 by the Homunculus SE lobe observed by Stahl et al. (2005) and derive, for the 
first time, the orbital and  Homunculus 
axis inclination. In Section 2 we 
describe the model and the 
adopted geometry for the system; in Section 3 we show the results, followed by the conclusions in Section 4.

\section{Model}

As the massive and supersonic winds of the two stars collide, they create a wind-wind 
interaction region, structured as two shocks 
at both sides of the contact discontinuity where the wind momenta are equal. 
As described in Luo, McCray \& Mac-Low (1990), the geometry of the contact discontinuity 
is asymptotically conical, with an opening angle $\beta$ given by:

\begin{equation}
\beta \simeq 120^{\circ} \left(1-\frac{\eta^{2/5}}{4} \right) \eta^{1/3},
\end{equation}

\noindent
where $\eta=\dot{M_s} v_s/\dot{M_p} v_p$, $\dot {M_p}$ and $\dot {M_s}$ are the mass loss rates and  $v_p$ and $v_s$ the wind velocities of the primary and secondary stars, respectively.

\begin{figure}
\centering
\includegraphics[width=8cm]{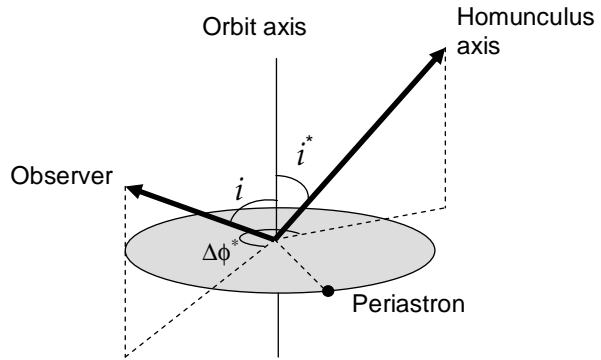}
\caption{Scheme of the system geometry. The orbital axis has an inclination $i$ 
regarding the line of sight 
and $i^*$ regarding the Homunculus axis. It is assumed that the angle between the line of 
sight and Homunculus axis is of $45^{\circ}$ (Davidson et al. 2001, Smith et al. 2004).}
\end{figure}

At the shock surfaces the compressed gas reaches temperatures of $\sim 10^6 - 10^8$K and emits free-free radiation, mostly at X-ray  wavelenghts. As this gas flows along 
the shock structure, it 
cools down to recombination temperatures and  becomes denser (Abraham et al. 2005a,b; 
Falceta-Gon\c calves, Jatenco-Pereira \& Abraham 2005). At this stage, it
could 
possibly be the source of He and H  lines, explaining some of the features observed in the spectra of massive 
binary systems (Falceta-Gon\c calves, Abraham \& 
Jatenco-Pereira 2006, 2007).

The model, proposed in Falceta-Gon\c calves et al. (2006), considers 
the emission from each subvolume of the conic shaped shock structure between the stars. 
The line intensity 
is obtained by the integration of each fluid element emission along the line of sight:

\begin{equation}
I(v) = \mathcal{C(\varphi)} \int_{0}^{\pi }\exp \left[ -\frac{\left( v-v_{\rm obs}\right) ^{2}}{2\sigma
^{2}}\right]e^{-\tau(\alpha)} d\alpha, 
\end{equation}

\noindent
where $v_{\rm obs} = v_{\rm flow} (-\cos \beta \cos \varphi\sin i+ 
\sin \beta \cos \alpha \sin \varphi \sin i - \sin \beta \sin \alpha \cos i)$,
is the observed stream velocity component projected into the line of sight, $\alpha$  the  azimuthal angle of 
the shock cone, $i$ the orbital inclination, $\varphi$ the orbital phase angle, $\sigma$ the turbulent 
velocity amplitude, $\tau$  the optical depth along the line of sight and $\mathcal{C(\varphi)}$  a normalization 
constant. 

From Eq. 2, it is possible to obtain  synthetic line profiles and compare them to the observations, as it was done with the HeII $\lambda4686$\AA \ emission line (Abraham \& Falceta-Gon\c calves 2007). To fit  the lines
reflected  by the Homunculus, it is also  necessary to take into account  the angle between the Homunculus axis and the line of sight $\Psi$, and the inclination of the 
 Homunculus axis with respect to the orbital plane.

In Fig. 1 we describe the geometry of the system as we assumed in the model. Here, $i$ and $i^*$
represent the inclination of the LOS and of the Homunculus axis with respect to the orbital axis, respectively. In this case, we may write the relationship between these angles as:

\begin{eqnarray}
\cos \Psi = \sin i \sin i^* \cos \Delta\phi^* + \cos i \cos i^* 
,\end{eqnarray}

\noindent
where $\Delta\phi^*$ represents  the angle between the projection of the Homunculus axis into the orbital plane and the line of sight.
Following 
Davidson et al.(2001) and Smith et al. (2004), we will consider the angle between the Homunculus axis and the LOS 
close 
to $\Psi \sim 45^{\circ}$.

\section{Observational data}

Stahl et al. (2005) presented VLT/UVES observations of the  line emission reflected by the SE 
lobe of the Homunculus, carried 
out from December 2002 to March 2004, or phase interval between 0.9 and 1.127, including the minimum event of 
2003.5.  The Paschen 
P8 line profile presented strong variability, with  clear P Cygni absorption, probably related to the stellar 
wind. Also, in all observations, two prominent excess emission bumps are visible, which we interpreted as extra emission from  the shock region. The digitalized excess 
emission velocity profiles are shown in Figure 2 as filled circles, corrected by
 the expansion velocity of the Homunculus $V = 100$ km s$^{-1}$.
  The numbers at each spectrum represent the date of the 
observation (JD - 2.400.000).

The observed peaks reflected by the Homunculus pole occurred at $v \sim 0$km s$^{-1}$ and $v \sim +220$km s$^{-1}$, while direct observations towards the central 
star revealed $v \sim 0$ and $v < 150$km 
s$^{-1}$ (Weis et al. 2005). This difference shows that, at least part of the observed emission 
has an anisotropic origin. We assumed that this latitude dependence is connected to the emission 
from the conic shock region.

\begin{figure*}
\centering
\includegraphics[width=6.2cm]{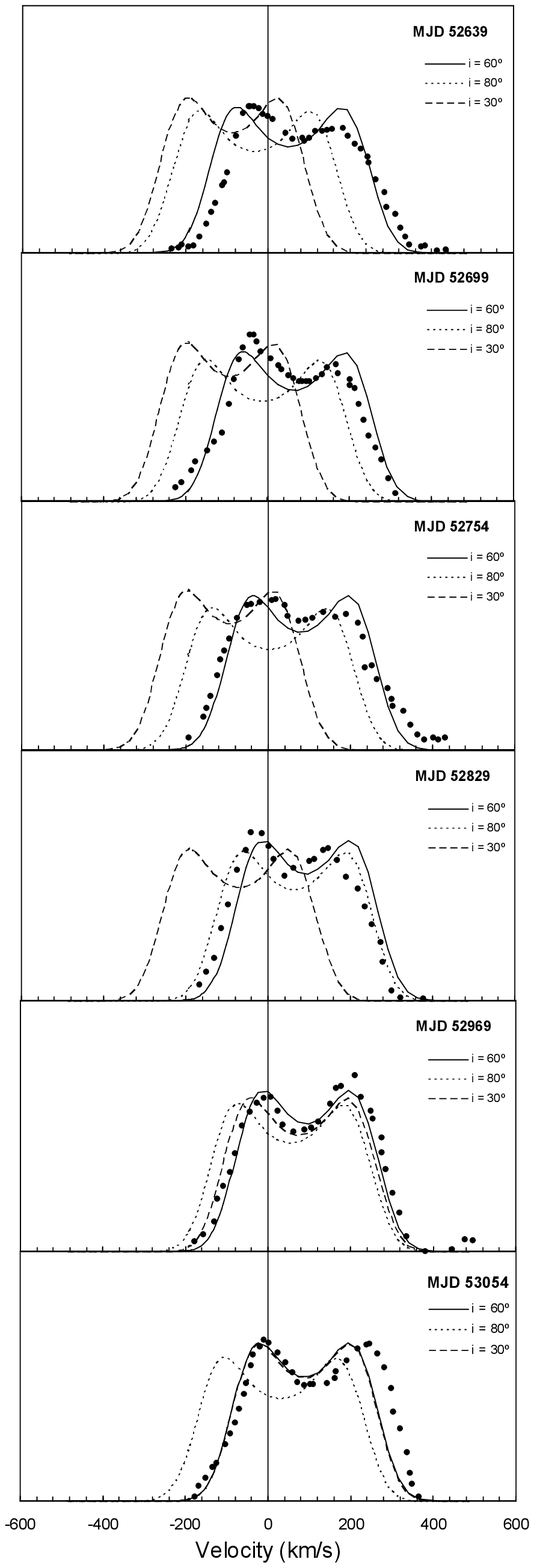}
\includegraphics[width=6.2cm]{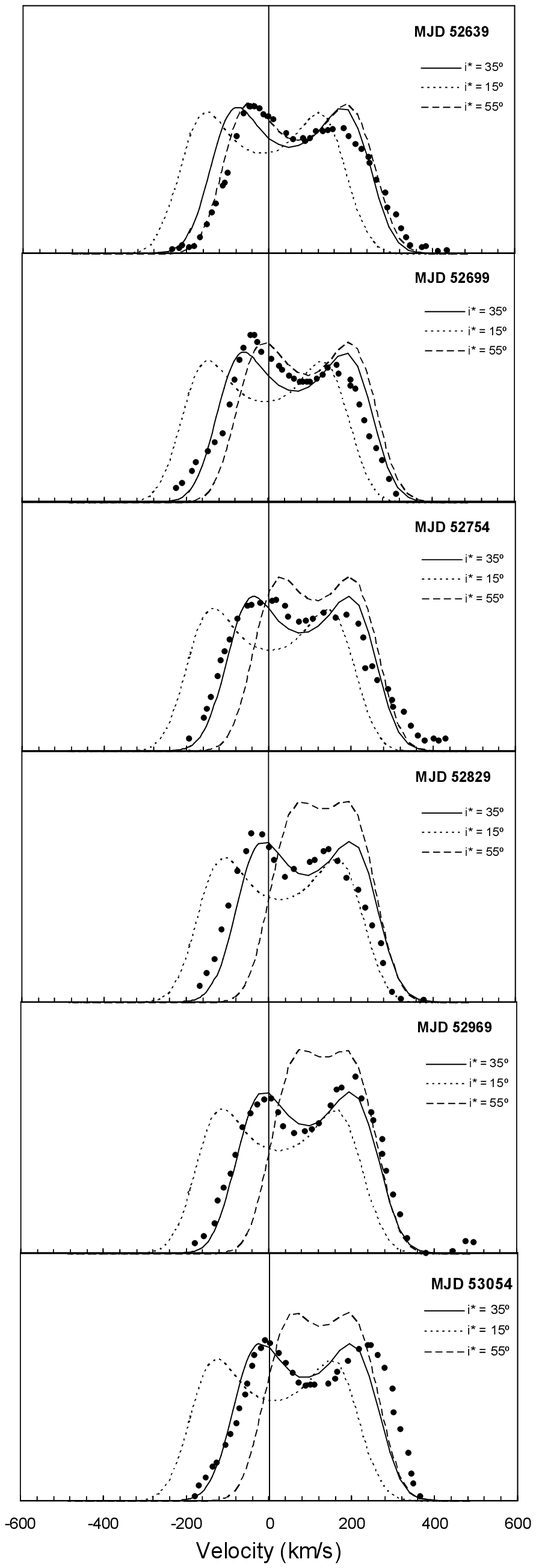}
\caption{Time series of the Paschen 8 line profiles, after continuum subtraction, 
measured for the Homunculus SE 
lobe. {\it Left:} the lines show the modeled line profiles for $i^* = 35^{\circ}$ and 
orbital inclinations regarding 
the line of sight $i=30^{\circ}$ (dashed), $60^{\circ}$ (solid) and $80^{\circ}$ 
(dotted). {\it Right:} the lines 
show the modeled line profiles for $i = 60^{\circ}$ and orbital inclinations regarding 
the Homunculus axis 
$i^*=55^{\circ}$ (dashed), $35^{\circ}$ (solid) and $15^{\circ}$ (dotted).}
\end{figure*}

\section{Results}

\begin{table}
\begin{center}
\caption{Parameters and input values}
\begin{tabular}{ccc}
\hline\hline 
parameter & input value & reference \\
\hline
$e$ & 0.95 & Abraham et al. (2005b) \\      
$\phi_0$ & -40$^\circ$ & Abraham \& Falceta-Gon\c calves (2007)\\ 
$P$  & 2024 days & Corcoran (2005) \\  
$a$ & 15A.U. & Falceta-Gon\c calves et al. (2005)\\ 
$t_0$ & 2,450,795 JD & Abraham et al. (2005b) \\
$\beta$ & 45$^{\circ}$ & Abraham \& Falceta-Gon\c calves (2007)\\ 
$v_{\rm flow} \sin\beta$ & 450 km s$^{-1}$ & Abraham \& Falceta-Gon\c calves (2007)\\
$\sigma$ & 0.33$v_{\rm flow}$ & Abraham \& Falceta-Gon\c calves (2007)\\
\hline\hline
\end{tabular}
\end{center}
\end{table}

We applied the model described in the  Section 2 to the observational data. For the wind 
parameters we used 
$\dot M_p = 2.5\times 10^{-3}$ M$_\odot$ yr$^{-1}$, $v_p = 700$ km s$^{-1}$,  
$\dot M_s = 5\times 10^{-5}$ M$_\odot$ yr$^{-1}$, $v_s = 3000$ km s$^{-1}$, which results
 in $\eta = 0.1$ and, from Eq. 1, we obtain $\beta=45^{\circ}$. For the flow along the conic surface we 
assumed $v_{\rm flow}\sin \beta =450$ km s$^{-1}$ and $\sigma = 0.33 v_{\rm flow}$ 
(Abraham \& Falceta-Gon\c calves 2007).
Because of the size of the Homunculus, we must also take into account the delay in 
the light travel from the wind-wind interaction zone to the reflection nebula. The 
data obtained from the reflection of the expanding nebula corresponds to an emission that 
occurred in a  phase previous than that observed directly towards the central star. This 
delay is particularly important for this model because a difference in the phase of the 
emission corresponds in a difference in the orientation of the shock cone regarding the 
Homunculus and, therefore, a difference shift velocity. The time delay may be calculated 
as:

\begin{equation}
\frac{t_{\rm delay}}{c} \sim \frac{t_{\rm expansion}}{v_{\rm expansion}},
\end{equation}

\noindent
considering an expansion of 100 km s$^{-1}$ since the outburst about 160 years ago, we get 
$t_{\rm delay} \sim 20$ days. Therefore, in order to precisely fit the model, we must 
subtract 20 days in the orbital 
phase and determine the actual orientation of the shock cone for the given phase angle. For the determination of the orbital phases we used the epoch of conjunction as June 29, 
2003 (Abraham et al. 2005b), 
$e=0.95$ and an angle between periastron and conjunction $\phi _0 = 40^{\circ}$ 
(Abraham \& Falceta-Gon\c calves 2007)

In Table 1 we show the input parameters of the model and their corresponding values. In 
the present work the input parameters, shown in Table 1, are fixed. We will 
briefly discuss the role of each of them on the obtained results later in the 
paper. Finally, we fitted the observational data using Eq. 2, together with Eq. 3, for 
an optically thin gas ($\tau \sim 0$). In Falceta-Gon\c calves et al. (2007) it has been 
shown that optically thick wind-wind shocks result in single peaked profiles, while the 
data of Paschen lines of $\eta$ Car presented double-peaked profiles at all phases. 
In order to obtain the best fitting set of 
values of $i$ and $i^*$, we calculate the synthetic line profiles for the 
observed epochs for the range $-180^{\circ}$ to $180^{\circ}$, independently for $i$ and 
$i^*$, in steps of $\Delta i = 10^{\circ}$. The observed data and three of the best 
models we fitted 
are shown in Fig. 2. The lines represent the models for different values of $i$ and 
$i^*$. It is noticeable from the modeled spectra that the line profiles are very 
sensitive to changes in the orbital inclination regarding both the observer and the 
Homunculus. The best fit, i.e. the minimum cumulative error for all orbital phases, is 
obtained for $i^* \sim 35^{\circ} \pm 10 ^{\circ}$ and $i \sim 60^{\circ} \pm 10 
^{\circ}$, i.e. an angle between the orbital plane and the line of 
sight of $\sim 30^{\circ}$, in agreement with the radio detection of the almost edge-on 
gas disk formed around the system. Note that the error of $\Delta i = 10^{\circ}$ 
represents 
the ``resolution" of the sets of $i$ and $i^*$ that we have calculated. Due to a small 
degeneracy between $i$ and $i^*$ for small values of $\Delta i$, it is not possible to 
reduce the imprecision in the determination of these parameters.

The most important issue regarding these results is that the Homunculus and the orbital 
plane axis are not aligned, as usually assumed (see Steiner \& Damineli 2004, Martin et al. 2006, Okazaki et al 2008). Since the Homunculus is assumed to be 
coincident with the 
stellar rotation axis at the epoch when the outburst event occurred, we may conclude that 
the stellar and orbital momenta were not parallel (or even close to that). Considering 
other massive binary systems, it is well known that the relaxation timescales for close 
binary systems are short, of order of hundred to a few thousands of years (e.g. Tassoul 
1990). {\it Should we expect the orbital and stellar angular momenta to coincide after 
the stellar evolution ($\sim 10^6$ yr) due to tidal relaxation?}

\subsection{Relaxation timescales}

Close massive binary systems are known to present strong tidal torques (Lecar, Wheeler \& McKee 
1976), which result in 
i-) orbital circularization, ii-) stellar rotation synchronization with the orbital period and iii-) 
alignment of stellar and orbital angular momenta. The last is particularly important 
considering the results obtained in this work.

To verify the plausibility of a non-aligned orbital and stellar momenta scenario, we may 
estimate the relaxing timescale of the $\eta$ Carinae system.
Several authors have addressed the problem is the past (e.g. Zahn 1977, Hut 1982, 
Eggleton, Kiseleva \& Hut 1998). The 
gravitational pull force acting on each star results in the non-spherical distribution of 
the stellar mass. However, since the stars are moving fast in close binaries, the 
orientation of the tides are not coincident with the direction of the centers of mass. The 
extra force component is responsible for the net torque that changes the energy and 
angular momentum of the orbital motion and stellar rotation. 

In particular, Hut (1982) studied the 
dynamical evolution of close binary systems with high eccentricity due to tidal friction 
between the two stars - a scenario we believe is correct for $\eta$ Carinae. In the case 
of high eccentricity, and a ratio of rotation to orbital angular momenta $\kappa << 
1$, the orbital and stellar angular momenta tend to become parallel for timescales of 
order:

\begin{equation}
\tau_{\rm align} \sim 0.02 T \left( \frac{D}{R_p} \right)^6 \left( 
\frac{r_g}{q} \right)^2 \frac{\Omega_p}{k \omega_p} \epsilon^{-3/2} (2-\epsilon)^{13/2},
\end{equation}

\noindent
where $k$ is the apsidal motion constant, $D$ is the distance between the stars at  
periastron passage, $q = M_s/M_p$ is 
the ratio between the stellar masses, $\epsilon = (1-e)$, $r_g = [I_p/(M_p 
R_p^2)]^{1/2}$,  
$I_p$ is the moment of inertia, $\Omega_p$ the angular velocity and $R_p$ the 
radius of the primary star,

\begin{equation}
\omega_p = [G (M_p + M_s) (1+e)]^{1/2} D^{-3/2},
\end{equation}

\noindent
is the orbital angular velocity at periastron, and:

\begin{equation}
T = \frac{R_p^3}{G M_p \tau},
\end{equation}

\noindent
where $\tau$ is the constant time lag of the tidal axis with respect to a coordinate 
system corotating with the primary star. The apsidal 
motion constant $k$ depends on the internal structure of the star, and is typically of 
the order of $10^{-3} - 10^{-1}$.

From Eqs. 5-7 it is clear that the relaxing timescale of the 
system depends on several poorly known parameters, though an estimation is still 
possible. For $D \sim 10 R_p$, $r_g/q \sim 1$, $M_s/M_p \sim 0.4$, 
$\Omega_p \sim \omega_p$ and $T \sim 10^8$s we have $\tau_{align} \sim 10^{16} - 
10^{18}$s, or $\sim 10^{8} - 10^{10}$yr, for $k = 10^{-3} - 10^{-1}$. From the parameters 
used in this estimation, the ratio $\Omega_p/\omega_p$ is the main unknown. Typically, we 
would expect a ratio $\sim 1$ for a relaxed system, but a ratio $> 1$ for a young wide 
binary system. In this case, the relaxation timescale would be even larger and our 
conclusions unchanged.

Typically, the lifetime of stars with $M > 60$M$_{\odot}$ is $1 - 5$ Myr (Schaller et al. 
1992). Therefore, we may conclude that $\eta$ Carinae system, during its evolution up to 
the outburst in the 19th century, had not enough time to relax, and we should not 
expect the orbital and rotation angular momenta to be aligned. Interestingly, this also 
reveals that the orbital high eccentricity ($e > 0.9$) may be primordial, even though it 
could also be a result of the massive mass ejection events.

\section{Conclusions}

Although many authors agree about the binary nature of $\eta$ Carinae, its orbital parameters and orientation are
still poorly known. From the X-ray and radio light curves, as well as the spectral line velocity curves, it is 
possible to derive  high eccentricity ($e > 0.9$) (Ishibashi et al. 1999, Abraham et 
al. 2005b, Falceta-Gon\c calves, Jatenco-Pereira \& Abraham 2005). Also, from the 
spectral line profiles it was believed that the periastron should happen in opposition to the observer, with 
$\phi_0 = 270^{\circ}$, if one assumes that the emission comes exclusively from the 
stellar winds (Davidson 1997, Nielsen et al. 2007). Abraham \& 
Falceta-Gon\c calves (2007) showed that the wind-wind shock 
region can be responsible for the HeII4686 emission 
excess, and consistently fit the observed spectra. However, in this case the periastron should occur near 
conjunction instead ($\phi_0 = 40^{\circ}$).

Regarding the orbital inclination, it is commonly assumed that the orbital plane should be nearly aligned with 
the Homunculus equator (Davidson 1997, Smith et al. 2004), which is inclined 
$\sim 45^{\circ}$ with respect to the 
observer. In contradiction to that, high resolution radio observations show the presence of a gas torus 
surrounding the system, which seems to be seen edge-on 
(Duncan \& White 2003), indicating that the orbit may not be aligned with the Homunculus. 
In this scenario, the 
torus would also be responsible for the shell-like effect seen in the X-ray light curve
(Falceta-Gon\c calves, Jatenco-Pereira \& Abraham 2005).

In this work we studied the possibility of the orbital plane and the Homunculus  not being aligned. For that, 
we used the hydrogen Paschen 8 line spectral profiles obtained from the SE lobe of the Homunculus. This line 
is supposed to originate in the central system and reflected by the expanding gas. 
The observed excess flux presented two peaked line profiles, which velocities change with the orbital phases. 
This behaviour is common in massive binary systems where the line emission originates in the wind-wind 
interaction zone. The conical shock surface would be resposible for an anomalous two peaked profile with 
velocities changing during the orbital period as 
the line of sight is intercepted by different vector projections of the flowing gas, as 
detailed in Falceta-Gon\c calves, Abraham \& Jatenco-Pereira (2006). 

We applied the synthetic line profile model, considering the emission from the central system, and derived the 
reflected fluxes at the SE lobe of the Homunculus. We fixed the stellar and orbital parameters $\beta = 45^{\circ}$, 
$e=0.95$, $v_{\rm flow}\sin \beta =450$ km s$^{-1}$, $\phi_0 = -40^{\circ}$ and $\sigma = 
0.33 v_{\rm flow}$ 
(Abraham \& Falceta-Gon\c calves 2007), and left the orbital inclination regarding the 
line of sight ($i$) and regarding the Homunculus 
axis ($i^*$) as fit parameters. We showed that the line profiles are very sensitive to changes in the orbital 
inclination regarding both the observer and the Homunculus axis. Therefore, this method has revealed helpful 
in constraining the orbital orientation of the system. The best fit was obtained for $i \sim 60 ^{\circ} \pm 10 ^{\circ}$ 
and $i^* \sim 35^{\circ} \pm 10 ^{\circ}$. This result shows that the line of sight must 
lie close to the orbital plane ($i \sim 30 ^{\circ}$), as indicated by radio observations 
and theoretical models regarding the X-ray light curve. 

In the presented calculations, most of the orbital and stellar parameters were fixed, 
based on previous references (Table 1). It is important though to understand if changes 
in any of these would reflect in major changes on our results and conclusions. 
The eccentricity of the orbit is a well-known 
parameter as it is constrained by the velocity curves of several spectral lines - we 
should expect its value to vary within the range $e = 0.9 - 0.95$. A different 
eccentricity would simply change the orbital phase, in 
which a given synthetic line would fit the data, therefore being irrelevant for the 
synthetic profiles. The same occurs for 
$\phi_0$, $P$, $a$ and $t_0$. Actually, $\phi_0$ is also related to the curvature of the 
shock cone. The turbulence amplitude $\sigma$, on the 
other hand, is responsible for the broadness of the synthetic line profiles. Since the 
observed broadness for each peak is 
unchanged at different orbital phases, its value is also irrelevant on the determination 
of $i$ and $i^*$. It is important though to remember that $\sigma$ presents a degeneracy 
with $v_{\rm flow}$, i.e. if the system has a smaller value of $v_{\rm flow}$, we would 
need a larger value of $\sigma$ to account for the same line broadness being $v_{\rm 
flow}$ directly obtained from the observed line velocities, but it has no influence on 
the determination of the inclination of the orbital plane. 
Among the fixed parameters, $\beta$ is the only strongly correlated with $i$ and $i^*$.  
As fully discussed in Falceta-Gon\c calves et al. (2006), a sequence of synthetic 
profiles may be well reproduced if an increased value of $\beta$ is balanced by a 
decrease of $i$, regarding the line of sight. However, considering the observed profiles, 
this degeneracy is limited 
to the range in which $i < \beta$ because, otherwise, models with $i > \beta$ result in 
single peaked profiles in certain orbital phases. Since the observational data presents 
two-peaked profiles in all phases we may constrain $i < \beta$. Fortunately, $\beta$ 
depends on the stellar winds of the two stars, which are approximately well constrained 
to provide 
$\beta \sim 40^{\circ} - 50^{\circ}$. This range of possible $\beta$ values is coherent 
with the obtained range of $i \sim 30 ^{\circ} \pm 10 ^{\circ}$, regarding the line of 
sight.

Also, the results show that the Homunculus axis is not perpendicular to 
the orbital plane. This suggests that the stellar angular momentum, probably coincident with the Homunculus axis, 
is not aligned with the orbital motion. We explained this fact showing that the relaxing timescale for momenta 
alignment and orbital circularization of $\eta$ Carinae is longer than the stellar 
lifetime. This fact also explains the high eccentricity of the system ($e \sim 0.95$), 
since tidal interactions did not have enough time to circularize the orbit. It is true 
though that the outburst occured in the 19th century could have been responsible for 
large changes in the orbital parameters as well. 

Finally, direct observations toward the central star revealed an increase in 
 P Cygni absorption during the minimum  (Davidson et al. 2005), as opposed  to the polar 
data, which shows almost constant absorption. This fact tell us that the main responsible 
for this absorption is not the primary star wind but, if the orbital plane is close to 
the LOS, the secondary star wind or the remaining post-shocked gas within the star and 
the observer. 

\section*{Acknowledgments}

D.F.G thanks FAPESP (No. 06/57824-1) and CNPq for financial support. Z.A. 
thanks FAPESP, CNPq and FINEP for support.

\end{document}